\documentclass[12pt]{article}

\usepackage{graphicx}
\begin{document}

\newcommand{\siml}{\stackrel{<}{\sim}}
\newcommand{\simg}{\stackrel{>}{\sim}}
\newcommand{\lleq}{\stackrel{<}{=}}
\newcommand{\vecx}{\mbox{\boldmath $x$}}

\baselineskip=1.333\baselineskip


%
\begin{center}
{\large\bf
Synchronization induced by periodic inputs 
in finite $N$-unit bistable Langevin models: 
The augmented moment method
} 
\end{center}

\begin{center}
Hideo Hasegawa
\footnote{E-mail address:  hideohasegawa@goo.jp
}
\end{center}

\begin{center}
{\it Department of Physics, Tokyo Gakugei University,  \\
Koganei, Tokyo 184-8501, Japan}
\end{center}
\begin{center}
({\today})
\end{center}
\thispagestyle{myheadings}

\begin{abstract}
We have studied the synchronization induced by periodic inputs
applied to the finite $N$-unit coupled bistable Langevin model
which is subjected to cross-correlated additive and multiplicative noises. 
Effects on the synchronization
of the system size ($N$), the coupling strength and
the cross-correlation between additive and multiplicative noise 
have been investigated with the use of 
the semi-analytical augmented moment method (AMM)
which is the second-order moment approximation for local
and global variables
[H. Hasegawa, Phys. Rev. E {\bf 67} (2003) 041903].
A linear analysis of the stationary solution of AMM equations shows that
the stability is improved (degraded) by positive (negative) couplings.
Results of the nonlinear bistable Langevin model are compared to
those of the linear Langevin model.

\end{abstract}

\vspace{0.5cm}





\newpage
\section{INTRODUCTION}

The Langevin model has been employed as 
a useful model for a wide range of stochastic phenomena
in physics, biology and chemistry 
(for reviews, see \cite{Gamma98,Lindner04}).
In the last decade, extensive studies have been made for
the bistable Langevin model given by
\cite{Wu94}-\cite{Kang08}
\begin{eqnarray}
\frac{dx_i}{dt} &=& x_i-x_i^3 +\frac{J}{N} \sum_{j=1}^N (x_j-x_i) +I(t)
+ \xi_i(t) + x_i \:\eta_i(t),
\hspace{1cm}\mbox{$i=1,...,N$},
\label{eq:A15}
\end{eqnarray}
where $J$ denotes the diffusive coupling, $I(t)$ an applied input,
and $\xi_i(t)$ and $x_i \eta_i(t)$ express 
additive and multiplicative noises, respectively
[for detail, see Eqs. (\ref{eq:A2b})-(\ref{eq:A2d})]. 
Refs. \cite{Wu94}-\cite{Zhang09} have studied
the bistable Langevin model for a single element with $N=1$ and $J=0$ 
in Eq. (\ref{eq:A15}). 
Because it is difficult to study the coupled bistable Langevin model 
given by Eq. (\ref{eq:A15}) with an {\it arbitrary} $N$, previous studies adopted
the mean-field model given by \cite{Desai78}-\cite{Kang08}
\begin{eqnarray}
\frac{dx}{dt} &=& (1-J) x -x^3 + J \bar{x}+I(t) + \xi(t)+ x\:\eta(x),
\label{eq:A16}
\end{eqnarray} 
which is  valid for $N=\infty$. 
Here $\bar{x}=\langle x \rangle$ represents the time-dependent order parameter,
and $\xi(t)$ and $x \eta(t)$ denote additive and multiplicative noises,
respectively. The phase transition and synchronization in coupled
bistable systems described by Eq. (\ref{eq:A16}) have been investigated 
\cite{Desai78}-\cite{Kang08}.

A common approach to {\it finite}-$N$ stochastic systems is to make 
direct simulations (DS)
for the Langevin equation given by Eq. (\ref{eq:A15}) 
or for relevant Fokker-Planck equation (FPE) \cite{Risken96}. 
Despite a recent development in computers, it is not easy to perform DEs of Eq. (\ref{eq:A15})
for appreciable values of $N$, which requires the computational time
growing as $N^2$ with increasing $N$.
For $N$-unit Langevin equations, 
the FPE method leads to $(N+1)$-dimensional
partial equations to be solved with proper
boundary conditions, which is usually very difficult.
As a useful semi-analytical method for stochastic equations, 
Rodriguez and Tuckwell \cite{Rod96}
proposed the moment method in which the first and second
moments of variables are taken into account.
In this approach, original $N$-dimensional
Langevin equations are transformed to $(N/2)(N+3)$-dimensional
deterministic equations. This figure becomes 65 and 5150
for $N=10$ and $N=100$, respectively. 
Based on a macroscopic point of view,
Hasegawa \cite{Hasegawa03a,Hasegawa06} has proposed 
the augmented moment method (AMM),
in which the dynamics of coupled Langevin
equations is described by a small number ({\it three}) of quantities for
averages and fluctuations of local and global variables.
The AMM has been  successfully applied to studies on the dynamics of 
coupled stochastic systems described by the linear Langevin model 
\cite{Hasegawa06,Hasegawa07}, 
FitzHugh-Nagumo model \cite{Hasegawa03a,Hasegawa08},
Hodgkin-Huxley model \cite{Hasegawa03b} and 
networks \cite{Hasegawa04c}. 
The AMM is a semi-analytical theory which is useful for a study 
of the finite-$N$ stochastic system, whereas conventional 
mean-field-type approximations are applied only to the $N=\infty$ case
\cite{Desai78}-\cite{Kang08}.
Cortical processing, for example, is performed by many
coupled populations of neurons and each population
consists of finite numbers of neurons.
Effects of cluster sizes on the information
processing in brain may be clarified in the AMM 
\cite{Hasegawa07b,Hasegawa09b}. 

Depending on the properties of elements forming a nonlinear system, 
we may classify them into two types A and B.
In the type A, elements are excitable units or
self-oscillators such as FitzHugh-Nagumo and Hodgkin-Huxley models. 
In the type B, units that form the system cannot oscillate
on their own unlike excitable systems. An example of the type B
is a bistable element, which has been
used in biology, chemistry and for a study of neural networks 
\cite{Sompolinsky86}-\cite{Koulakov02}. 
Interesting and intrigue properties of the bistable
Langevin model have been extensively investigated 
\cite{Wu94}-\cite{Kang08}.
Stationary probability distribution, first-passage time and
the stochastic resonance for subthreshold periodic inputs
in the bistable Langevin model subjected to cross-correlated 
additive and multiplicative white (or colored) noise
have been studied.
There are, however, still many unsolved basic problems.
For example, dynamics of the finite-$N$ bistable Langevin model
when time-dependent inputs are applied has not been well understood.
The purpose of the present paper is to apply the AMM
to the bistable Langevin model.
We will investigate the dynamical response to external periodic inputs in 
the finite-$N$ bistable Langevin model
subjected to additive and multiplicative noises with the use of the AMM.
We apply {\it suprathreshold} inputs:
deriving inputs are suprathreshold in the sense that
they can induce the emergence of the transition of states from
one of the bistable states to the other.
We may investigate the synchronization of forced oscillations 
induced by suprathreshold periodical inputs in the type-B system, just as in the case of
the type-A ensemble where the synchronization among self-oscillations are studied.
We can examine effects on the synchronization of the size of systems,
coupling strength and the cross-correlation between
additive and multiplicative noises 
in a semi-analytical way within the AMM.
The synchronization in the finite-$N$ coupled bistable model
is recently investigated by numerical methods \cite{Casado07}.

The paper is organized as follows.
In Sec. 2, the AMM is applied
to coupled bistable Langevin model subjected to additive and multiplicative 
noise with the cross-correlation.
Numerical model calculations are presented in Sec. 3.
In Sec. 4, we make a linear stability analysis of the stationary solution.
The dynamical properties of the bistable Langevin
model are compared to those of the linear Langevin model.
Sec. 5 is devoted to our conclusion.

\section{Augmented moment method}
\subsection{Bistable Langevin model}

Generalizing the model given by Eq. (\ref{eq:A15}),
we adopt the $N$-unit coupled bistable
Langevin model given by
\begin{eqnarray}
\frac{dx_i}{dt}\!\!&=&\!\!F(x_i)+ \xi_i(t) + G(x_i) \:\eta_i(t)
+I_i^{(c)}(t)+I^{(e)}(t), 
\label{eq:A1}
\end{eqnarray}
with 
\begin{equation}
I_i^{(c)}(t)=\frac{J}{Z} 
\sum_{k(\neq i)} [x_k(t)-x_i(t)]
\hspace{1.0cm}\mbox{($i=1,...,N$)}.
\label{eq:A2} 
\end{equation}
Here $F(x)=- \partial U(x)/\partial x$, $U(x)$ denotes the potential: 
$G(x)$ is an arbitrary function of $x$:
$J$ expresses the diffusive coupling:
$Z$ (with $Z=N-1$) stands for the coordination number: 
$I^{(e)}(t)$ is an external input:
$\xi_i(t)$ and $\eta_i(t)$ express zero-mean Gaussian white
noises with correlations given by
\begin{eqnarray}
\langle \eta_i(t)\:\eta_j(t') \rangle
&=& \alpha^2 \:\delta_{ij} \delta(t-t'),
\label{eq:A2b}\\
\langle \xi_i(t)\:\xi_j(t') \rangle 
&=& \beta^2 \:\delta_{ij} \delta(t-t'),
\label{eq:A2c}\\
\langle \eta_i(t)\:\xi_j(t') \rangle 
&=& \epsilon \: \alpha \beta \: \delta_{ij} \delta(t-t'),
\label{eq:A2d}
\end{eqnarray}
where $\alpha$ and $\beta$ denote the strengths of multiplicative
and additive noises, respectively, and
$\epsilon$ the cross-correlation between additive and multiplicative
noises. 

We will study the dynamical properties 
of the coupled Langevin model
with the use of the AMM \cite{Hasegawa03a,Hasegawa06}, 
in which the three quantities of
$\mu$, $\gamma$ and $\rho$ are defined by
\begin{eqnarray}
\mu(t) &=& \langle X(t) \rangle 
= \frac{1}{N} \sum_i \langle x_i(t) \rangle, \\
\label{eq:A4} 
\gamma(t) &=& \frac{1}{N} \sum_i 
\langle [x_i(t)-\mu(t)]^2 \rangle, \label{eq:A5} \\
%
\rho(t) &=& \langle [X(t)-\mu(t)]^2 \rangle.
\label{eq:A6}
\end{eqnarray}
Here $X(t)$ with $X(t)=N^{-1}\sum_i x_i(t)$ expresses a global variable,
$\mu$ its mean, and $\gamma$ and $\rho$ denote
fluctuations in local ($x_i$) and global ($X$) variables,
respectively.
Equations of motion for $\mu$, $\gamma$ and $\rho$
are given by (the argument $t$ being suppressed:
for details, see appendix A)
\begin{eqnarray}
\frac{d \mu}{dt}&=& f_0+f_2 \gamma + 3 f_4 \gamma^2
+\frac{\phi}{2}\{ \alpha^2 [g_0g_1+3(g_1g_2+g_0g_3)\gamma]
+ \epsilon \alpha \beta(g_1+3g_3 \gamma)\} +I^{(e)}, \nonumber \\
&&
\label{eq:A7} \\
\frac{d \gamma}{dt} &=& 2f_1 \gamma + 6 f_3 \gamma^2
+ (\phi+1) (g_1^2+2 g_0g_2)\alpha^2\gamma 
+ 2 \phi \epsilon \alpha \beta g_2 \gamma 
+ \left( \frac{2 J N}{Z} \right)(\rho-\gamma) + P, \nonumber \\
&& 
\label{eq:A8} \\ 
%
\frac{d \rho}{dt} &=& 2 f_1 \rho 
+ 6 f_3 \gamma \rho 
+(\phi+1)(g_1^2+2 g_0g_2)\alpha^2\rho 
+ 2 \phi \epsilon \alpha \beta g_2 \rho 
+ \frac{P}{N},
\label{eq:A9}
\end{eqnarray}
with
\begin{eqnarray}
P &=& \alpha^2 g_0^2+2 \epsilon \alpha \beta (g_0+g_2 \gamma) +\beta^2,
\label{eq:A10}
\end{eqnarray}
where $f_{\ell}=(1/\ell !)
\partial^{\ell} F(\mu)/\partial x^{\ell}$, 
$g_{\ell}=(1/\ell !) 
\partial^{\ell} G(\mu)/\partial x^{\ell}$,
and $\phi=0$ and 1 in the Ito and Stratonovich representations,
respectively.
The $O(\gamma^2)$-order terms in Eqs. (\ref{eq:A7})-(\ref{eq:A9}) 
are included with the use of the Gaussian approximation given by 
\cite{Hasegawa03a}\cite{Tanabe01}
\begin{eqnarray}
\langle (\delta x_i)^3 \rangle &\simeq& 0, 
\label{eq:A11} \\
\langle (\delta x_i)^4 \rangle &\simeq& 
3 \langle (\delta x_i)^2 \rangle^2, 
\label{eq:A12} \\
\langle (\delta x_i)^2(\delta x_j)^2 \rangle &\simeq& 
3 \langle (\delta x_i)^2 \rangle  \langle (\delta x_j)^2 \rangle.
\label{eq:A13}
\end{eqnarray}
These terms 
play crucial roles for the bistable Langevin model
although they are not necessary for the linear Langevin model 
\cite{Hasegawa06}.
Original $N$-dimensional stochastic equations 
given by Eqs. (\ref{eq:A1}) and (\ref{eq:A2})
are transformed to three-dimensional deterministic equations
given by Eqs. (\ref{eq:A7})-(\ref{eq:A10}). 

For the bistable Langevin model with 
\begin{eqnarray}
F(x) &=& x-x^3,\\
G(x) &=& x,
\end{eqnarray}
equations of motion are given by 
\begin{eqnarray}
\frac{d \mu}{dt}&=& \mu-\mu^3 - 3 \mu \gamma 
+ \frac{\alpha^2 \mu}{2}+\frac{\epsilon \alpha \beta}{2}+I(t),
\label{eq:B1} \\
\frac{d \gamma}{dt} &=& 2 (1 - 3 \mu^2 - 3 \gamma) \gamma
+ 2 \alpha^2 \gamma + \left( \frac{2 J N}{Z} \right)(\rho-\gamma)  
+ P, 
\label{eq:B2} \\
\frac{d \rho}{dt} &=& 2(1- 3 \mu^2 - 3 \gamma) \rho
+ 2 \alpha^2 \rho + \frac{P}{N},
\label{eq:B3}
\end{eqnarray}
with
\begin{eqnarray}
P &=& \alpha^2 \mu^2 + 2 \epsilon \alpha \beta \mu + \beta^2.
\label{eq:B4}
\end{eqnarray}
Eqs. (\ref{eq:B1}) and (\ref{eq:B2}) with $N=1$ ($J=\epsilon=0$) 
are in agreement with results
obtained for a single bistable Langevin model \cite{Zhang06}, while
those with $N=\infty$ ($\rho=0$) agree with results
obtained for $N=\infty$ bistable Langevin model 
with a mean-field approximation \cite{Kang08}.
For $J=0$, Eqs. (\ref{eq:B2}) and (\ref{eq:B3}) lead to
\begin{eqnarray}
\rho &=& \frac{\gamma}{N},
\label{eq:B5}
\end{eqnarray}
which expresses the central-limit theorem.

A linear stability analysis of the stationary solution will be
made in Sec. 4.1 with the use of the deterministic AMM equations given by
Eqs. (\ref{eq:B1})-(\ref{eq:B3}),
as in the case of the coupled FitzHugh-Nagumo model \cite{Hasegawa08}.

\subsection{Synchrony}


In order to quantitatively study the emergence 
of a synchronized state of the ensembles defined by Eqs. (\ref{eq:A1}) and (\ref{eq:A2}),
we first consider the quantity $S'(t)$ given by
\begin{equation}
S'(t)=\frac{1}{N^2} \sum_{i j}<[x_{i}(t)-x_{j}(t)]^2>
=2 [\gamma(t)-\rho(t)].
\label{eq:C1}
\end{equation}
When all variables are in the same state: $x_{i}(t)=X(t)$ for all $i$
(the completely synchronous state), we obtain $S'(t)=0$ in Eq. (\ref{eq:C1}).
On the contrary, in the asynchronous state where $\rho=\gamma/N$,
it is given by $S'(t)=2(1-1/N)\gamma(t) \equiv S'_{0}(t)$
\cite{Hasegawa03a}.
We may define the normalized ratio for the synchrony given by 
\cite{Hasegawa03a}
\begin{equation}
S(t) \equiv 1-\frac{S'(t)}{S'_{0}(t)}
= \left( \frac{N}{Z}\right) 
\left[\frac{\rho(t)}{\gamma(t)}-\frac{1}{N} \right],
\label{eq:C2}
\end{equation}
which is 0 and 1 for completely asynchronous ($S'=S'_0$) and synchronous 
states ($S'=0$), respectively.

We may alternatively interpret $S(t)$ as the normalized
mutual correlation given by
\begin{eqnarray}
S(t) &=& \frac{\zeta(t)}{\gamma(t)},
\label{eq:C3}
\end{eqnarray}
with
\begin{eqnarray}
\zeta(t) &=& \frac{1}{NZ} \sum_i \sum_{j(\neq i)}
[x_i(t)-\mu(t)][x_j(t)-\mu(t)], \\
&=& \left( \frac{N}{Z} \right)
\left[ \rho(t)-\frac{\gamma(t)}{N} \right].
\label{eq:C4}
\end{eqnarray}
We note that $S(t)=0$ for $J=0$ [Eq. (\ref{eq:B5})].

The equation of motion for $S(t)$ is given by
\begin{eqnarray}
\frac{d S}{d t} &=& - \left( \frac{P}{\gamma} \right) S
+ \frac{2 J}{Z}(ZS+1)(1-S),
\end{eqnarray}
with the use of Eqs. (\ref{eq:B2}), (\ref{eq:B3}) and (\ref{eq:C2}).

\section{Model Calculations}

We have performed model calculations,
solving the AMM equations by the Runge-Kutta method with a time step of 0.01.
Direct simulations for the $N$-unit Langevin model
have been performed by using the Box-Mueller algorithm and
the Euler method \cite{Sancho82}-\cite{Mei06} with a time step of 0.0001.
Results are averages of 1000 trials. 

Our model given by Eqs. (\ref{eq:A1}) and (\ref{eq:A2}) includes five parameters of
$N$, $J$, $\alpha$, $\beta$ and $\epsilon$.
We will investigate effects of $N$, $J$ and $\epsilon$
for fixed values of $\alpha=0.1$ and $\beta=0.1$ otherwise noticed.

\vspace{0.5cm}
\noindent
{\bf Periodic pulse inputs}

We apply periodic pulse inputs at $t \geq t_1$ given by
\begin{eqnarray}
I^{(e)}(t)
&=& A\: \sum_k [\Theta(t-t_1- kT_p) \Theta(t_1+ kT_p+t_w-t) 
\nonumber \\
&-& \Theta(t-t_1-k T_p/2) \Theta(t_1+k T_p/2 +t_w-t)],
\label{eq:D1}
\end{eqnarray}
with $A=1.0$, $t_1=50$, $T_p=100$ and $t_w=10$
where $\Theta(x)$ denotes the Heaviside function:
$\Theta(x)=1$ for $x \geq 0$ and zero otherwise.
Figure \ref{fig1}(a), (b) and (c) show the time courses of
$\mu(t)$, $\gamma(t)$ and $S(t)$, respectively, for
$N=10$, $J=0.2$ and $\epsilon=0.5$.
Initial values of $x_i(t=0)$ are set to be $-1.0$.
When a positive pulse input $I(t)$ shown by the chain curve
in Fig. \ref{fig1}(a) is applied at $t = 50$
to the state which has been randomized by noise,
the average value of $\mu$ is changed to about $1.0$.
When a negative pulse input is applied to the state with
$\mu\simeq 1.0$ at $t=100$, the state is switched back 
to $\mu\simeq -1.0$.
In these switching process, the local fluctuation $\gamma(t)$ and
synchronization $S(t)$ are transiently increased.
In order to investigate the relation among $\mu$, $\gamma$ and $S$,
$\gamma$ and $S$ calculated by the AMM are plotted 
as a function of $\mu$ in Fig. \ref{fig1}(d). It is shown
that in the process of $\mu(t)$ changing from $-1.0$ to $+1.0$,
$S(t)$ has the maximum value at $\mu \simeq 1.0$.  
In the reversed process, $S(t)$ has a maximum value 
at $\mu \simeq -1.0$. 
The maximum value of $S(t)$ for a process from $\mu\simeq -1.0$
to $\mu \simeq 1.0$ is smaller than that for the inversed process,
which is due to the introduced cross-correlation ($\epsilon=0.5$)
between additive and multiplicative noise.

Similar $\mu-S$ plots for positive and negative correlations are
depicted in Figs. \ref{fig2}(a) and (b), respectively.
Figure \ref{fig2}(a) shows that with increasing $\epsilon$,
the maximum value at $\mu\simeq 1.0$ ($\mu\simeq -1.0$) is 
increased (decreased).
The reversed behavior is realized for negative $\epsilon$,
as shown in Fig. \ref{fig2}(b):
the $\mu-S$ plot for negative $\epsilon$ is symmetric 
to that for positive $\epsilon$.


Figures \ref{fig3}(a) and (b) show the $J$ dependent $\mu-S$ plots
for positive and negative $J$, respectively.
From a comparison between Figs. \ref{fig3}(a) and (b),
we note that the positive coupling is more effective than
the negative one in increasing the synchrony: note that the
vertical scale of Fig. \ref{fig3}(b) is smaller than that of
Fig. \ref{fig3}(a).


The maximum value of $S$, $S_{max}$, 
is plotted as a function of $N$ in Fig. \ref{fig4}
which shows that the synchrony is more increased for larger $J$
and smaller $N$. 

\vspace{0.5cm}
\noindent
{\bf Sinusoidal inputs}

Next we apply sinusoidal inputs at $t \geq t_1$ given by
\begin{eqnarray}
I^{(e)}(t)
&=& A\: \sin \left( \frac{2 \pi t}{T_p}\right)
\Theta(t-t_1),
\label{eq:D2}
\end{eqnarray}
with $A=1.0$, $t_1=50$ and $T_p=100$.
Figures \ref{fig5}(a), (b) and (c) show time courses of
$\mu$, $\gamma$ and $S$, respectively, and
Fig. \ref{fig5}(d) shows the relevant $\mu-S$ plot.
From a comparison between Figs. \ref{fig1} and \ref{fig5},
we note that the magnitudes of $\gamma$ and $S$ for the
sinusoidal input are about three times larger than that for pulse input.
The $\mu-S$ plots for periodic pulse and sinusoidal inputs are similar besides
their magnitudes.

\section{Discussion}
\subsection{Stability analysis}

We will investigate the stability of the stationary
solution of Eqs. (\ref{eq:B1})-(\ref{eq:B3}), from which
the Jacobian matrix is given by
\begin{eqnarray}
\left(
\begin{array}{ccc}
{\displaystyle 1-3 \mu^2-3\gamma+ \frac{\alpha^2}{2} }
& -3 \mu & 0 \\
{\displaystyle -12 \mu\gamma +2(\alpha^2 \mu+\epsilon \alpha \beta)}
&{\displaystyle 2(1-3\mu^2-6\gamma+\alpha^2) -\frac{2JN}{Z} } 
& {\displaystyle \frac{2JN}{Z} }\\
{\displaystyle -12\mu\rho +\frac{2}{N}(\alpha^2 \mu+\epsilon \alpha \beta)}
& -6\rho
& 2(1-3\mu^2-3\gamma+\alpha^2)
\end{array}
\right). \nonumber \\
&& \label{eq:Y1}
\end{eqnarray}

In the case of $I=J=\epsilon=0$, we may analytically
obtain stationary solutions and eigenvalues of the Jacobian matrix. 
Stationary solutions are divided into two cases $A$ and $B$
in which $\mu=0$ and $\mu \neq 0$, respectively:
the latter is further classified to cases $B_1$ and $B_2$ as follows.

\noindent
(1) Case $A$

Stationary solutions are given by
\begin{eqnarray}
\mu^2&=&0, \\
\gamma&=& \frac{1}{6}\left( 1+\alpha^2+\sqrt{D_1} \right), \\
\rho&=& \frac{1}{6 N}\left( 1+\alpha^2+\sqrt{D_1} \right),
\end{eqnarray}
and relevant eigenvalues are given by
\begin{eqnarray}
\lambda_1 &=& \frac{1}{2}\left[1-\sqrt{D_1} \right], \\
\lambda_2 &=& -2 \sqrt{D_1}, \\
\lambda_3 &=& 1+\alpha^2-\sqrt{D_1}, 
\end{eqnarray}
with
\begin{eqnarray}
D_1 &=& (1+\alpha^2)^2+6 \beta^2.
\label{eq:Y2}
\end{eqnarray}

\noindent
(2) Case $B_1$

Stationary solutions are given by
\begin{eqnarray}
\mu^2&=&\frac{1}{2}(1 + \sqrt{D_2}), \\
\gamma&=& \frac{1}{6}\left( 1+\alpha^2 - \sqrt{D_2} \right), \\
\rho&=& \frac{1}{6N}\left( 1+\alpha^2 - \sqrt{D_2} \right),
\end{eqnarray}
and relevant eigenvalues are given by
\begin{eqnarray}
\lambda_{1,2} &=& -2 -\sqrt{D_2} \pm \sqrt{4-3 D_2},\\
\lambda_3 &=& -2+\alpha^2 -2 \sqrt{D_2},
\end{eqnarray}
with
\begin{eqnarray}
D_2 &=& 1-\alpha^2-\frac{\alpha^4}{2}-3 \beta^2.
\label{eq:Y3}
\end{eqnarray}

\noindent
(3) Case $B_2$

Stationary solutions are given by
\begin{eqnarray}
\mu^2&=&\frac{1}{2}(1 - \sqrt{D_2}), \\
\gamma&=& \frac{1}{6}\left( 1+\alpha^2 + \sqrt{D_2} \right), \\
\rho&=& \frac{1}{6 N}\left( 1+\alpha^2 + \sqrt{D_2} \right),
\end{eqnarray}
and relevant eigenvalues are given by
\begin{eqnarray}
\lambda_{1,2} &=& -2 +\sqrt{D_2} \pm \sqrt{4-3 D_2},\\
\lambda_3 &=& -2+\alpha^2 +2 \sqrt{D_2}.
\end{eqnarray}
When $\alpha$ and $\beta$ are small, stationary solutions and
eigenvalues are expressed in powers of $\alpha$ and $\beta$: 
those in the lowest approximation are summarized in Table 1.
We realize that the solution of $A$ is stable while that of
$B_2$ is unstable because its first eigenvalue of $\lambda_1$ 
becomes positive.
The solution of $B_1$ is stable if $Re(\lambda_1) < 0$ (see below).

\renewcommand{\arraystretch}{1.5}
\begin{table}
\begin{center}
\caption{Stationary solutions and eigenvalues 
within $O(\alpha^2)$ and $O(\beta^2)$
}
\begin{tabular}{|c||c|c|c|} \hline
\hspace{2cm} & Case $A$ & Case $B_1$ & Case $B_2$ \\ \hline \hline
$\mu^2$ &  0  &  $\frac{1}{4}(4-\alpha^2-3\beta^2)$ 
& $\frac{1}{4}(\alpha^2+3\beta^2)$ \\ \hline
$\gamma$ & $\frac{1}{6}(2+2\alpha^2+3\beta^2)$  
&  $\frac{1}{4}(\alpha^2+\beta^2)$ 
& $\frac{1}{12}(4+\alpha^2-3\beta^2)$ \\ \hline 
$\rho$ & $\frac{1}{6N}(2+2\alpha^2+3\beta^2)$  
&  $\frac{1}{4N}(\alpha^2+\beta^2)$ 
& $\frac{1}{12N}(4+\alpha^2-3\beta^2)$ \\ \hline \hline
$\lambda_1$ & $-\frac{1}{2}(\alpha^2+3 \beta^2)$  
&  $-2+2\alpha^2+6 \beta^2$ & $\alpha^2+3 \beta^2$ \\ \hline
$\lambda_2$ & $-2(1+\alpha^2+3 \beta^2)$  
&  $-4-\alpha^2-3 \beta^2$ & $-2-2\alpha^2-6\beta^2$ \\ \hline
$\lambda_3$ & $-3 \beta^2$  
&  $-4+2\alpha^2+3 \beta^2$ & $-3 \beta^2$ \\ \hline
\end{tabular}
\end{center}
\end{table}

It has been shown that although results of the stability condition derived by
the second-order moment method are not in good agreement with those of
numerical method, it yields semi-quantitatively 
meaningful results \cite{Kang08}.
Bearing this fact in mind, we will study the stability condition
for the case of $\mu \sim \pm 1$ (case $B_1$) with 
$I=\epsilon=0$ within the AMM by numerical methods.
The solid curve in Fig. \ref{fig6} shows the calculated boundary
for $J=0.0$
within which the stationary solution is stable
(results for $J \neq 0.0$ in Fig. \ref{fig6} will be explained shortly).
Figure \ref{fig7}(a) shows the $\alpha$ dependence of
the maximum eigenvalues, $\lambda_{max}$, for $\beta=0.0$ and $N=10$.
For $J=0.0$, $\lambda_{max}$ becomes zero for $\alpha=0.855$ ($\beta=0.0$),
above which the stationary solution becomes unstable.
When the coupling of $J$ is introduced, the critical
$\alpha$ value becomes 0.738, 0.968 and 1.106 for $J=-0.2$,
0.2 and 0.5, respectively.
Figure \ref{fig7}(b) shows a similar plot of the $\beta$
dependence of $\lambda_{max}$ for $\alpha=0.0$ and $N=10$.
The solution becomes unstable for $\beta > 0.577$
with $J=0.0$, $\alpha=0.0$ and $N=10$. With introducing $J$,
the critical $\beta$ value is changed to 0.518, 0.633 and 0.712
for $J=-0.2$, 0.2 and $J=0.5$, respectively.
Similar calculations of $\lambda_{max}$ are made for
$\alpha=0.5$ with changing $J$.
Squares, circles and triangles in Fig. \ref{fig6}
express results calculated for $J=0.5$, 0.2 and -0.2, respectively,
which are shown by curves for a guide of eye. 
Figure \ref{fig6} shows that the stability of the stationary solution 
against additive and multiplicative
noise is improved (degraded) by positive (negative) couplings.

\subsection{Effect of the symmetry of $G(x)$}
It is necessary to point out that the symmetry in the $\mu-S$ plot
depends on the symmetry of $G(x)$ for multiplicative noise.
Indeed, in the case of $G(x)=x$
which has the odd symmetry: $G(x)=-G(-x)$,
the asymmetry in the $\mu-S$ plot is obtained as shown in Figs. \ref{fig2}(a) and (b).
However, if $G(x)$ has the even symmetry: $G(x) =G(-x)$,
the asymmetry in the $\mu-S$ plot is not realized.
We have performed the AMM calculation for 
$G(x)= x^2-1$ with the even symmetry, for which equations of motion
for $\mu$, $\gamma$ and $\rho$ are given in appendix C.  
Calculated $S(t)$ and $\gamma(t)$ are plotted against $\mu(t)$ in
Fig. \ref{fig8}(a) for $\epsilon \geq 0$ and in Fig. \ref{fig8}(b) 
for $\epsilon \leq 0$. 
With changing $\epsilon$, $\mu-S$ plot is little modified being
symmetric independently of $\epsilon$, 
although magnitudes of the $\mu-\gamma$ plot is changed. 
A comparison of Figs. \ref{fig8}(a) and (b) with Figs. \ref{fig2}(a) and (b)
clearly shows that the symmetry of $G(x)$ is important
in studying the effect of the cross-correlation between additive
and multiplicative noise.
This fact may be applied to
effects of the cross-correlation on the 
stationary probability distribution as shown below.
From the FPE in Eq. (\ref{eq:X1}), the stationary distribution $p(\vecx)$ for
$I=J=0$ in the Stratonovich representation ($\phi=1$) is expressed by
\begin{eqnarray}
p(\vecx) &=& \prod_{i=1}^N \:p(x_i),
\end{eqnarray}
with
\begin{eqnarray}
\ln p(x) &\sim & \int 
\frac{2 F(x)}{[\alpha^2 G(x)^2+2 \epsilon \alpha \beta G(x)+\beta^2]}\:dx 
- \left(\frac{1}{2}\right) 
\ln\left[\alpha^2 G(x)^2+2 \epsilon \alpha \beta G(x)+\beta^2 \right].
\nonumber \\ &&
\end{eqnarray}
Straightforward calculations for $F(x)=x-x^3$, $G(x)=x$ and $G(x)=x^2-1$ lead to
\begin{eqnarray}
p(x) &\propto & 
(\alpha^2 x^2 + 2 \epsilon \alpha \beta x +\beta^2)
^{[\alpha^2+\beta^2(1-4 \epsilon^2)]/\alpha^4-1/2} 
\nonumber \\
&& \times \exp\left[-\frac{\alpha^2 x^2-4 \epsilon \alpha \beta x}{\alpha^4}
-\frac{2 \epsilon (\alpha^2+\beta^2(3-4 \epsilon^2)) }{\alpha^4 \sqrt{1-\epsilon^2}}
\tan^{-1}\left(\frac{\epsilon \beta + \alpha x}{\beta \sqrt{1-\epsilon^2}} \right) \right] 
\nonumber \\
&&\hspace{9cm}\mbox{for $G(x)=x$}, \\
p(x) &\propto&
[\alpha^2 (x^2-1)^2 + 2 \epsilon \alpha \beta(x^2-1)+\beta^2 ]^{-(1/2\alpha^2+1/2)} 
\nonumber \\
&& \times \exp\left[\frac{\epsilon }{ \alpha^2 \sqrt{1-\epsilon^2}}
\tan^{-1} \left(\frac{\epsilon \beta+ \alpha (x^2-1) }
{\beta \sqrt{1-\epsilon^2}}\right) \right] 
\hspace{1cm}\mbox{for $G(x)=x^2-1$}. 
\end{eqnarray}
Figures \ref{fig9}(a) and (b) show the stationary distributions for
$G(x)=x$ and $G(x)=x^2-1$, respectively, with $\alpha=\beta=0.5$
for various $\epsilon$.
For $G(x)=x$, an introduction of $\epsilon$ yields the asymmetry
in $p(x)$, and $p(x)$ for a negative $\epsilon$ is anti-symmetric
with that for a positive $\epsilon$ with respect to the $x=0$ axis. 
In contrast, for $G(x)=x^2-1$, $p(x)$ is symmetric independently
of $\epsilon$, and an effect of a negative $\epsilon$ is
different from that of a positive $\epsilon$.
The difference in $G(x)$ reflects on various aspects
of the bistable Langevin model
such as the stationary distribution, the mean first-passage time 
and stochastic resonance, which have been conventionally
calculated with the use of $G(x)=x$
\cite{Wu94}-\cite{Zhang09}.

\subsection{The linear Langevin model}
It is worthwhile to compare the properties of the nonlinear bistable Lanvevin
model to those of the linear Langevin model.
For the linear Langevin model with $F(x)=- \kappa x$ 
($\kappa$: relaxation rate) and $G(x)=x$,
we obtain equations of motion for $\mu$, $\gamma$ 
and $\rho$ in the Stratonovich representation 
as given by
\begin{eqnarray}
\frac{d \mu}{dt}&=&-\kappa \mu 
+ \frac{\alpha^2 \mu}{2}+\frac{\epsilon \alpha \beta}{2}+I(t), 
\label{eq:E1}\\
\frac{d \gamma}{dt} &=& -2 \kappa \gamma 
+ 2 \alpha^2 \gamma + \left( \frac{2 J N}{Z} \right)(\rho-\gamma)  
+ P, 
\label{eq:E2}\\
\frac{d \rho}{dt} &=& -2\kappa \rho
+ 2 \alpha^2 \rho + \frac{P}{N},
\label{eq:E3}
\end{eqnarray}
with
\begin{eqnarray}
P &=& \alpha^2 \mu^2 + 2 \epsilon \alpha \beta \mu + \beta^2.
\label{eq:E7}
\end{eqnarray}
Equations (\ref{eq:E1})-(\ref{eq:E7}) with $\epsilon=0$ agree 
with those previously obtained \cite{Hasegawa06}.
We may obtain analytic expressions for the stationary state,
given by
\begin{eqnarray}
\mu &=& \frac{2 I-\epsilon \alpha \beta}{2 \kappa -\alpha^2}, 
\label{eq:E4} \\
\gamma &=& \frac{P}{2(\kappa-\alpha^2-JN/Z)}
\left[1 + \frac{J}{Z(\kappa -\alpha^2)} \right], 
\label{eq:E5} \\
\rho &=& \frac{P}{2 N (\kappa-\alpha^2)},
\label{eq:E6}
\end{eqnarray}
yielding
\begin{eqnarray}
S &=& \frac{J}{J+Z(\lambda-\alpha^2)},
\end{eqnarray}
where $P$ is given by Eq. (\ref{eq:E7}) with $\mu$ in Eq. (\ref{eq:E4}).

Figure \ref{fig10} (a),(b) and (c) show time courses of 
$\mu$, $\gamma$ and $S$, respectively, for the linear Langevin
model calculated with $\kappa=1.0$ and the same parameters 
for $N$, $J$, $\alpha$, $\beta$ and $\epsilon$ as in Fig. \ref{fig1}.
The relevant $\mu-S$ plot is depicted in Fig. \ref{fig10}(d).
A comparison of Fig. \ref{fig10} with Fig. \ref{fig1}
shows that dynamical behavior of the bistable Langevin
model is quite different from those of the linear counterpart. 
The nonlinearity in the bistable Langevin model plays an important role 
for the synchronization in the ensemble given by Eqs. (\ref{eq:A1}) and (\ref{eq:A2}).

We may make a linear analysis of the stationary solution given 
by Eqs. (\ref{eq:E4})-(\ref{eq:E6}).
From Eqs. (\ref{eq:E1})-(\ref{eq:E3}), we obtain the Jacobian matrix given by
\begin{eqnarray}
\left(
\begin{array}{ccc}
{\displaystyle -\kappa+\frac{\alpha^2}{2} }
& 0 & 0 \\
{\displaystyle 2 \alpha^2 \mu}
&{\displaystyle -2\kappa+2 \alpha^2-\frac{2JN}{Z} } 
& {\displaystyle \frac{2JN}{Z} }\\
{\displaystyle \frac{2\alpha^2 \mu}{N}}
& 0& -2 \kappa+2\alpha^2
\end{array}
\right). 
\end{eqnarray}
Eigenvalues are given by
\begin{eqnarray}
\lambda_1 &=& -\kappa+\frac{\alpha^2}{2}, \\
\lambda_2 &=& -2 \kappa+2 \alpha^2-\frac{2JN}{Z}, \\
\lambda_3 &=& -2 \kappa+2 \alpha^2.
\end{eqnarray}
We note that eigenvalues in the linear Langevin model are independent 
of $I$ and $\beta$, which is different from those 
in the nonlinear bistable Langevin model.

\section{Conclusion}
We have studied the synchronization induced by periodic pulse and sinusoidal inputs
in the $N$-unit bistable Langevin model subjected to 
cross-correlated additive and multiplicative noise 
with the use of the semi-analytical AMM \cite{Hasegawa03a,Hasegawa06}.
It has been shown that 

\noindent 
(1) the synchrony is transiently increased when the mean value of
state variables is switched from one stable state to the other which is induced
by an external suprathreshold input,

\noindent
(2) the magnitude of synchrony is increased with increasing 
the coupling strength ($J$) and/or decreasing
the system size ($N$),

\noindent
(3) The stability of the stationary solution against additive and/or multiplicative
noise is improved by positive couplings but degraded by
negative couplings, 

\noindent
(4) the effect of the cross-correlation depends on its 
symmetry as well as a functional form of $G(x)$ for the multiplicative noise,
and

\noindent
(5) properties of the nonlinear bistable Langevin model are rather different from
those of the linear Langevin model.

\noindent
The AMM \cite{Hasegawa03a,Hasegawa06} may be 
applied not only to the type-A stochastic ensembles 
of excitable elements but also
to the type-B ones consisting of nonexcitable elements. 
It is expected possible to apply the AMM to various types of coupled stochastic 
systems. In the AMM we may easily solve the three-dimensional deterministic equations 
of $\mu$, $\gamma$ and $\rho$ for periodic as well as  
non-periodic (transient) inputs, 
although its applicability is limited to the small-noise case
which is inherent in the moment method.

\section*{Acknowledgements}
This work is partly supported by
a Grant-in-Aid for Scientific Research from the Japanese 
Ministry of Education, Culture, Sports, Science and Technology.  

\vspace{0.5cm}
\appendix

\noindent
{\large\bf Appendix A: Derivation of the AMM equations}
\renewcommand{\theequation}{A\arabic{equation}}
\setcounter{equation}{0}

The Fokker-Planck equation for the Langevin model given by 
Eqs. (\ref{eq:A1}) and (\ref{eq:A2})
is given by 
\cite{Hasegawa06}
\begin{eqnarray}
\frac{\partial}{\partial t}\: p(\vecx,t) 
&=&-\sum_k \frac{\partial}{\partial x_i} \left( \left [F(x_i) + I_i
+\frac{\phi}{2}[\alpha^2 G'(x_i)G(x_i)+\epsilon \alpha \beta G'(x_i)]
\right ] \:p(\vecx,t) \right)  
\nonumber \\
&+&\frac{1}{2}\sum_{i }\frac{\partial^2}{\partial x_i^2} 
\left( [\alpha^2 G(x_i)^2 + 2 \epsilon \alpha \beta G(x_i)
+\beta^2]\:p(\vecx,t) \right),
\label{eq:X1}
\end{eqnarray}
where $p(\vecx,t)$ ($\vecx= \{ x_k \}$), 
$I_k=I_k^{(c)}+I$, $G'(x)=dG(x)/dx$, and
$\phi=1$ and 0 in the Stratonovich and Ito representations,
respectively.

With the use of Eqs. (\ref{eq:X1}), equations of motion
are given by \cite{Hasegawa06}
\begin{eqnarray}
\frac{d \langle x_i \rangle}{dt}
&=& \langle F(x_i) \rangle
+\langle I_i \rangle
+\frac{\phi \:\alpha^2}{2} \langle G'(x_i)G(x_i) \rangle
+ \frac{\phi \epsilon \alpha \beta}{2} \langle G'(x_i) \rangle, 
\label{eq:X2} \\
\frac{d \langle x_i \:x_j \rangle}{dt}
&=& \langle x_i\:F(x_j) \rangle 
+ \langle x_j\: F(x_i) \rangle 
+ \langle x_i I_j \rangle + \langle x_j I_i \rangle
\nonumber \\
&+& \frac{\phi\:\alpha^2}{2}
[\langle x_i G'(x_j) G(x_j) \rangle
+ \langle x_j G'(x_i) G(x_i)\rangle] \nonumber \\ 
&+& \frac{\phi \epsilon \alpha \beta}{2}
[\langle x_i G'(x_j) \rangle
+ \langle x_j G'(x_i) \rangle] \nonumber \\ 
&+&[\alpha^2\:\langle G(x_i)^2 \rangle 
+ 2 \epsilon \alpha \beta \langle G(x_i) \rangle 
+\beta^2]\:\delta_{ij},
\label{eq:X3} \\
\frac{d \langle X \rangle}{dt}
&=&\frac{1}{N} \sum_i \frac{d \langle x_i \rangle}{dt}, 
\label{eq:X4} \\
\frac{d \langle X^2 \rangle}{dt} 
&=& \frac{1}{N^2}\sum_i \sum_j 
\frac{d \langle x_i\:x_j \rangle}{dt},
\label{eq:X5}
\end{eqnarray}
where 
$X=N^{-1} \sum_i x_i$.
Expanding $x_i$ in Eqs. (\ref{eq:X2})-(\ref{eq:X5}) 
around the average value of $\mu$ as
\begin{equation}
x_i=\mu+\delta x_i,
\end{equation}
we obtain equations of motion for $\mu$, $\gamma$ and $\rho$
given by Eq. (\ref{eq:A7})-(\ref{eq:A10})
with the Gaussian approximation [Eqs. (\ref{eq:A11})-(\ref{eq:A13})].

\vspace{0.5cm}
\noindent
{\large\bf Appendix B: The AMM equations for $G(x)=x^2-1$}
\renewcommand{\theequation}{B\arabic{equation}}
\setcounter{equation}{0}

For the bistable Langevin model with 
\begin{eqnarray}
F(x) &=& x-x^3,\\
G(x) &=& x^2-1,
\end{eqnarray}
Eqs. (\ref{eq:A7})-(\ref{eq:A10}) yield equations of motion given by 
\begin{eqnarray}
\frac{d \mu}{dt}&=& \mu-\mu^3 - 3 \mu \gamma 
+ \alpha^2 \mu (\mu^2-1+3 \gamma)
+\epsilon \alpha \beta+I(t),
\label{eq:Z1} \\
\frac{d \gamma}{dt} &=& 2 (1 - 3 \mu^2 - 3 \gamma) \gamma
+ 4 \alpha^2 \gamma (3 \mu^2-1) + 2 \epsilon \alpha \beta \gamma
+ \left( \frac{2 J N}{Z} \right)(\rho-\gamma)  
+ P, 
\label{eq:Z2} \\
\frac{d \rho}{dt} &=& 2(1- 3 \mu^2 - 3 \gamma) \rho
+ 4 \alpha^2 \rho (3 \mu^2-1) + 2 \epsilon \alpha \beta \rho
+ \frac{P}{N},
\label{eq:Z3}
\end{eqnarray}
with
\begin{eqnarray}
P &=& \alpha^2 (\mu^2-1)^2 + 2 \epsilon \alpha \beta (\mu^2-1+\gamma) 
+ \beta^2.
\label{eq:Z4}
\end{eqnarray}

\newpage


\newpage

\begin{figure}
\begin{center}
\end{center}
\caption{
(Color online)
Time courses of (a) $\mu$, (b) $\gamma$ and (c) $S$,
and (d) $\gamma$ and $S$ as a function of $\mu$ for the pulse input 
with $N=10$, $J=0.2$, $\alpha=0.1$, $\beta=0.1$ and $\epsilon=0.5$.
Solid and dashed curves an (a)-(c) denote results of AMM and DS, respectively:
the chain curve in (a) shows input $I(t)$ shifted downward by - 2.0:
arrows in (d) express the direction of time development.
}
\label{fig1}
\end{figure}

\begin{figure}
\begin{center}
\end{center}
\caption{
(Color online)
$\epsilon$ dependence of the $\mu-S$ plot:
(a) for $\epsilon=0.0$ (the solid curve), $\epsilon=0.5$ (the dashed curve)
and $\epsilon=0.9$ (the chain curve),
and (b) for $\epsilon=0.0$ (the solid curve), $\epsilon=-0.5$ (the dashed curve)
and $\epsilon=-0.9$ (the chain curve)
with $N=10$ and $J=0.2$.
}
\label{fig2}
\end{figure}

\begin{figure}
\begin{center}
\end{center}
\caption{
(Color online)
$J$ dependence of the $\mu-S$ plot:
(a) The $\mu-S$ plot 
for $J=0.1$ (the dashed curve), $J=0.2$ (the solid curve)
and $J=0.5$ (the chain curve), and
(b) for $J=-0.1$ (the dashed curve), $J=-0.2$ (the solid curve)
and $J=-0.5$ (the chain curve) with $N=10$ and $\epsilon=0.0$.
}
\label{fig3}
\end{figure}

\begin{figure}
\begin{center}
\end{center}
\caption{
(Color online)
The maximum value of synchrony $S_{max}$ against $N$ 
for $(J, \epsilon)=(0.5, 0.0)$ (the solid curve), $(0.2,0.0)$ (the dashed curve), and
$(0.2,0.5)$ (the chain curve) with $\alpha=0.1$ and $\beta=0.1$.
}
\label{fig4}
\end{figure}

\begin{figure}
\begin{center}
\end{center}
\caption{
(Color online)
Time courses of (a) $\mu(t)$, (b) $\gamma(t)$ and (c) $S(t)$,
and (d) $\gamma$ (the chain curve) and $S$ (the solid curve)
as a function of $\mu$ for the sinusoidal input
with $N=10$, $J=0.2$, $\alpha=0.1$, $\beta=0.1$ and $\epsilon=0.5$.
Solid and dashed curves an (a)-(c) denote results of AMM and DS, respectively:
the chain curve in (a) shows input $I(t)$ shifted 
downward by - 2.0: arrows in (d) express the direction of time development.
}
\label{fig5}
\end{figure}

\begin{figure}
\begin{center}
\end{center}
\caption{
(Color online)
The $\alpha$-$\beta$ phase boundary 
for $J=0.5$ (the chain curve), 0.2 (the dotted curve),
0.0 (the solid curve) and $-0.2$ (the dashed curve) 
calculated within the AMM with $N=10$:
the result for $J=0.0$ is calculated by $D_2 =0$ 
[Eq. (\ref{eq:Y3})], and marks (filled circles, squares and triangles) 
for $J \neq 0.0$ show numerical results with curves plotted 
for a guide of eye (see text). 
}
\label{fig6}
\end{figure}

\begin{figure}
\begin{center}
\end{center}
\caption{
(Color online)
(a) The maximum eigenvalue, $\lambda_{max}$, as a function of 
$\alpha$ with $\beta=0.0$, and
(b) $\lambda_{max}$ as a function of 
$\beta$ with $\alpha=0.0$ 
for $J=-0.2$ (dashed curves), 0.0 (solid curves),
0.2 (dotted curves) and $0.5$ (chain curves) 
with $N=10$ (see text).
}
\label{fig7}
\end{figure}

\begin{figure}
\begin{center}
\end{center}
\caption{
(Color online)
$S$ and $\gamma$ versus $\mu$ for pulse input 
with $G(x) = x^2-1 $,
(a) $S$ for $\epsilon=0.0$ (the solid curve) 
and $\epsilon=0.5$ (the dashed curve):
$\gamma$ for $\epsilon=0.0$ (the chain curve) 
and $\epsilon=0.5$ (the dotted curve);
(b) $S$ for $\epsilon=0.0$ (the solid curve) 
and $\epsilon=-0.5$ (the dashed curve):
$\gamma$ for $\epsilon=0.0$ (the chain curve) 
and $\epsilon=-0.5$ (the dotted curve)
($N=10$ and $J=0.2$) (see text).
}
\label{fig8}
\end{figure}

\begin{figure}
\begin{center}
\end{center}
\caption{
(Color online)
The stationary distribution $p(x)$ for (a) $G(x)=x$
and (b) $G(x)=x^2-1$ for various values of $\epsilon$
with $\alpha=\beta=0.5$, the ordinate of (a) being different from
that of (b).
}
\label{fig9}
\end{figure}

\begin{figure}
\begin{center}
\end{center}
\caption{
(Color online)
Time courses of (a) $\mu$, (b) $\gamma$ and (c) $S$,
and (d) the $\mu-S$ plot for the pulse input 
applied to the linear Langevin model
with $N=10$, $J=0.2$, $\lambda=1.0$, $\alpha=0.1$, $\beta=0.1$ and
$\epsilon=0.5$.
Solid and dashed curves an (a)-(c) denote results of AMM and DS, respectively:
the chain curve in (a) shows input $I(t)$ shifted downward 
by - 2.0: arrows in (d) express the direction of time development.
}
\label{fig10}
\end{figure}

\end{document}